# Tunable acoustic energy concentrations based on pseudo-spin locking waveguides and topological rainbow trappings

Bowei Wu, Teng Wang, Shuanghuizhi Li, Tingfeng Ma[*]

Zhejiang-Italy Joint Lab for Smart Materials and Advanced Structures, School of Mechanical Engineering & Mechanics, Ningbo University, Ningbo, 315211, China

## ABSTRACT

In this work, tunable acoustic energy concentrations are realized based on pseudo-spin locking waveguides and topological rainbow trappings. Firstly, a tunable pseudo-spin locking is proposed, and the broad acoustic energy transport and spin-locked one-way transport are verified. The results show that acoustic wave transports based on pseudo-spin locking waveguides are more robust to structure defects than conventional topological edge-state waveguides. Besides, topological rainbow trappings are realized by adjusting the distribution of liquid in tubes. Based on those, we investigate the coupling of the pseudo-spin locking waveguides and the topological rainbow trappings. The results show that high acoustic energy concentrations can be obtained conveniently by using the coupling energy concentrator based on the pseudo-spin locking waveguides and the topological rainbow trappings. The results present a novel method to concentrate acoustic wave energy, which is vital in the application field of acoustic sensings and microfludics.

The according experimental verifications will be presented in the near future.

*matingfeng@nbu.edu.cn

# 1. Introduction

In recent years, topological insulators have attracted significant attentions in the fields of optics[1–3], acoustics[4–6], and mechanics[7–12], which are immunity to backward scatterings and manufacturing defects, thus ensure robust energy transmissions. A variety of topological transport states based on quantum Hall effects [13,14], quantum valley Hall effects [15–17], and quantum spin Hall effects [18–21] have been proposed and realized. These topological states have been used to realized reconfigurable waveguides[22], beam splitters[23], topological switches[24] etc. However, most topological interface states confine energy to a fixed narrow interface, which significantly restricts the transmission of high-capacity energy, which limits its applicability.

Recently, a kind of broad waveguide energy transmission has been proposed in recent years [25–29], which consists of three distinct layers. This configuration involves a Dirac point-band structure positioned between two domains with different topological phases[30–32]. This specific arrangement facilitates the creation of a topological waveguide structure with an adjustable width [33,34]. Some researchers have demonstrated the transmission properties of valley-locked systems based on valley Hall effects [35,36]. On the other hand, He et al. [37]proposed a new method to create a broad topological waveguide based on a photonic quantum spin Hall system. They experimentally observed the unique property of unidirectional propagation due to pseudo-spin momentum locking. Chen et al. [38]have also proposed broad topological acoustic waveguide states based on spin hall effect.

Most researches on broad-waveguide energy transmissions are composed of passive materials, resulting in a fixed waveguide path and energy concentration degree [33], which poses limitations for application scenarios requiring energy accumulations with different degrees, such as acoustic wave sensings and acousticfluidics. Besides, acoustic-wave energy accumulations and capturing have been realized by using topological rainbow trapping[39–42], however, the energy concentration degrees of which are also not tunable. Moreover, the waveguides of topological rainbow trappings are based on interface states, the path of which is not broad enough. In recent years, in biomedical field, broad waveguides with tunable energy concentration degree and path become necessary in integrated sensings and particle manipulations. Thus, it is essential to propose a method to realize a flexible broad topological waveguide with tunable energy concentration degree and path.

In this work, pseudo-spin locking waveguides with tunable energy concentration degree and path has been investigated. Based on that, the tunable energy concentrator based on the coupling of pseudo-spin locking waveguides and the topological rainbow trapping has been realized. Crucially, it is found that the coupled system can easily amplify the energy concentration by nearly 17 times. Moreover, due to the tunability of the system, the amplification and gathering position of the energy can also be adjusted flexibly. This research will bring new research ideas for energy aggregation and large-channel acoustic wave sensings in the future.

## 2. Model

The structure of the model is shown in Fig.1. The hexagonal unit cell is shown in Fig.1(b), the substrate of which contains 18 holes. The substrate material is acrylic. The holes are filled with different levels of water. By varying the water height in these holes, different unit cells can be obtained. For example, at R=31.5mm, let $h_d \neq 0$, at R= 15.5 and 24 mm, set $h_s = 0$, this unit cell is marked as Type-A; At R=24mm, let $h_d \neq 0$, at R= 31.5 and 15.5 mm, set $h_s = 0$, this unit cell is marked as Type-B; At R=15.5mm, let $h_d \neq 0$, at R=24 and 31.5 mm, set $h_s = 0$, this unit cell is marked as Type-C. In Fig.1(a), an acrylic plate is placed on the substrate, the position of which controls the height of the air wall. In this design, adjusting the height of the air wall also leads to changes of the frequency range.

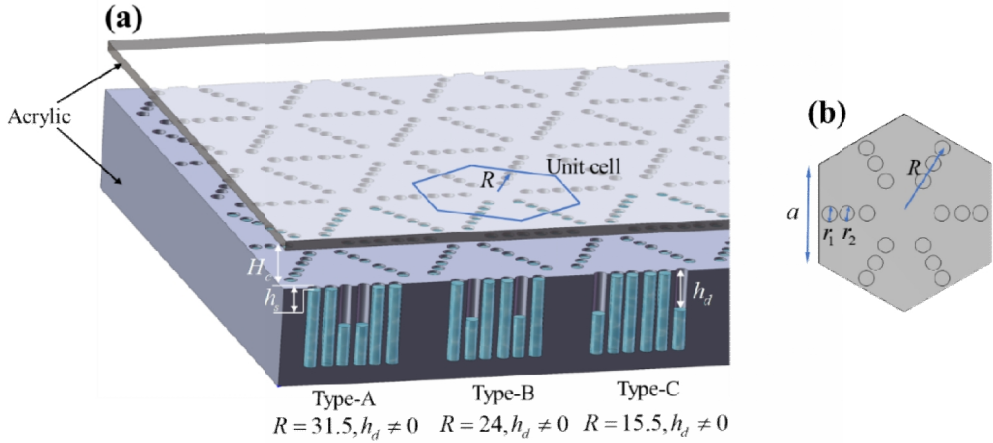

Fig.1. (a) The schematic diagram of topological phononic crystal composed of an acrylic ceiling and an acrylic substrate. The distance between the ceiling and the substrate is $H_c$, and the cylindrical cavities of the phonon crystals are distributed in the hexagonal lattice. (b) The unit cell, it consists of 12 cavities with a radius of $r_1 = 3.2$ mm and 6 cavities with a radius of $r_2 = 3.1$ mm, with a lattice constant of $a = 24$ mm. The effective depths $h_s$ and $h_d$ of the cavities can be varied by pumping water into the cavity. When $R = 31.5$ mm, $h_d \neq 0$, it is marked as a Type-A unit cell; when R = 24 mm, $h_d \neq 0$, it is marked as Type-B unit cell; and when R = 15.5 mm, $h_d \neq 0$, it is marked as a Type-C unit cell.

The pressure acoustics module of COMSOL Multiphysics is used to calculate the acoustic pressure distribution. The acoustic transmission medium is set as air (denstiy is $\rho = 1.2 \text{kg/m}^3$, sound velocity is $c = 343 \text{m/s}$). Periodic boundary conditions are imposed on the opposite edges of the hexagonal unit cell, heights of the air columns are $h_d = 20$ mm and $h_s=0$, and the distance between the acrylic plate and the substrate is $H_C = 15$ mm. For type-A unit cell, $R = 31.5$ mm and $h_d = 20$ mm; For type-B unit cell, $R = 24$ mm and $h_d = 20$ mm; For type-C unit cell, $R = 15$ mm and $h_d = 20$ mm. Due to the band folding mechanism, the Dirac point folds into a double Dirac point at M-point, as illustrated in Fig. 2(c) and (d). It is possible to create phononic crystals with topologically non-trivial states by extending the distance of the six cylinders from the center (i.e., the air columns appear at a distance of R = 31.5 mm). Due to the $C_{6v}$ symmetry of the unit cell, the $\{p_x, p_y\}$ and $\{d_{xy}, d_{x^2-y^2}\}$ orbitals form double degenerate pairs at the M-point. Alternatively, the distance of the six cylinders from the center is reduced (i.e., the air columns appear at a distance of R = 15.5 mm) to create phononic crystals with trivial states, shown in Fig. 2(e) and (f). This contraction can lead to the band inversion, which demonstrates the structures in both cases exhibiting different topological properties, as shown in Fig. 2(g).

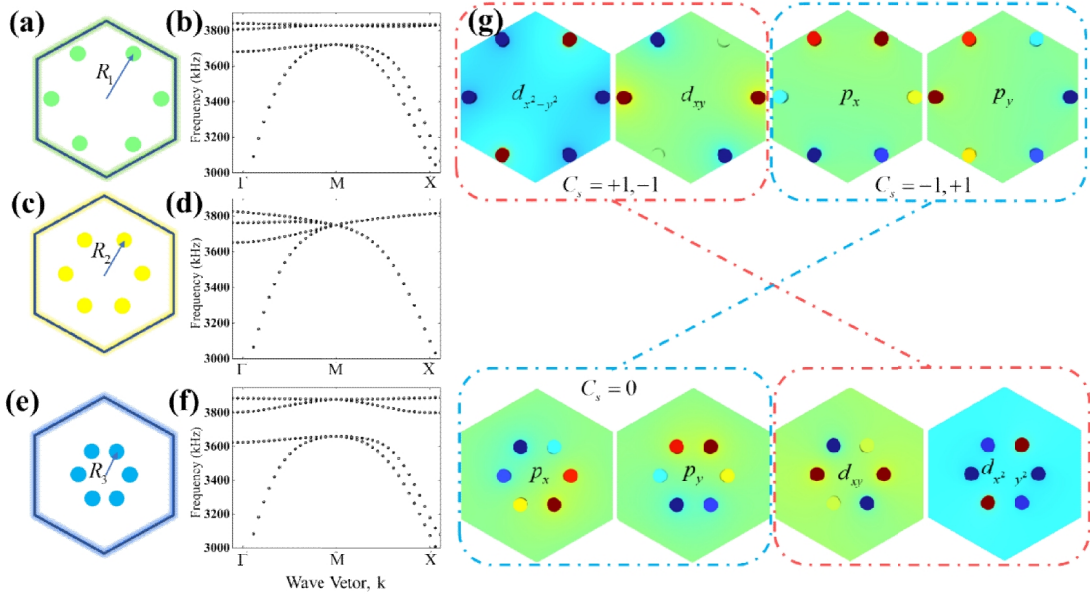

Fig.2 Diagrams of three types of unit cells. (a) R = 31.5 mm, Type-A unit cell. (b) The dispersion curves of Type-A unit cell. (c) R = 24 mm, Type-B unit cell. (d) The dispersion curves of Type-B unit cell (e) R = 15.5 mm, Type-C unit cell. (f) The dispersion curves of Type-C unit cell. (g) Acoustic pressure of several unit cells. When R decreases from 31.5mm to 15.5mm, the resulting band map evolution shows that $\{p_x, p_y\}$ and $\{d_{xy}, d_{x^2-y^2}\}$ models undergo band inversion. Corresponds to a topological transition from a topological insulator (TI, non-zero $C_s$) to an ordinary insulator (OI, zero spin Chen number ($C_s$)).

In order to verify the topological properties, we calculate the spin chern number of the two structures to confirm the topological phase transition. According to our calculations, the spin chern number ($C_s$) value of the expansion unit cell in Fig 2(a) is non-zero ($\pm 1$), indicating a topological property. In contrast, the contraction unit cell in Fig 2(e) has a $C_s$ value of 0, indicating a trivial property. An artificial spin $\pm\frac{1}{2}$ can be formed through both configurations.

A sandwich structure of phonon crystal is constructed, namely, a Type-B phononic crystal with a double Dirac cone is placed between a Type-A phononic crystal and a Type-C phononic crystal, a helical waveguide with a broad pseudo-spin locking can be formed.

## 3. Broad pseudo-spin locking waveguide

To demonstrate that a broad pseudo-spin locking waveguide can be created, we simulate acoustic transmission properties of a three-layer heterostructure consisted of phononic crystals with types A, B, and C, respectively. In the simulations, periodic and radiative boundary conditions are applied to the $x$ and $y$ directions of the supercells, where Type-A, B and C have 4, 2, and 4 unit cells, respectively, and the band structure is shown in Fig. 3(b). It is shown that a pair of edge states (two red lines) emerge, which exhibit the helical characteristics of quantum spin Hall edge states, characterized by two modes with opposite group velocities at each frequency. To achieve a pseudo-spin locking waveguide, the intermediate Dirac cone domain need be sandwiched between two domains with different topological properties, according to the body-edge correspondence principle. Due to the coupling of the Type A-B and Type B-C interface states, the intermediate domain B is characterized by both the topological interface states of the A-B and B-C interfaces, and the bulk state of the intermediate domain B. To demonstrate that the relevant waveguide modes depicted in Fig. 3(b) possess a large scale, we selected two representative modes, indicated by red and blue circles, field distributions of which are presented on the right side of Fig. 3(b). It is evident that the energy amplitudes are almost uniformly distributed in the central region. Besides, there exist other non-topological waveguide modes that emerge within the bandgap, as indicated by the blue dots in Fig. 3(b).

Then, the numbers of unit cell in Type-A and Type-C phononic crystals are fixed to be four. The number of unit cell of in the middle field Type-B is increased, noted by Type-$B_x$. Fig.2(c) illustrates the topological band vs. frequency for the intermediate domain width $x$. As $x$ increases, the size of the topological waveguide frequency (green region) decreases due to weakened couplings between the A|B and B|C domain wall interface modes and the intermediate B domain bulk modes. The supercell bands for each intermediate domain with width $x$ are shown in Figs. 2(d)-(j). Notably, at $x = 0$ and 1, the blue dot representing the non-topological mode is absent. However, as the value of $x$ increases, the non-topological modes (blue dots) progressively shift into the bandgap from the bulk states (black dots). This is because the couplings between the A|B and B|C domain-wall interface modes becomes weaker with increasing $x$, resulting that more non-topological modes (blue dots) gradually compress the operating range of the topological waveguide.

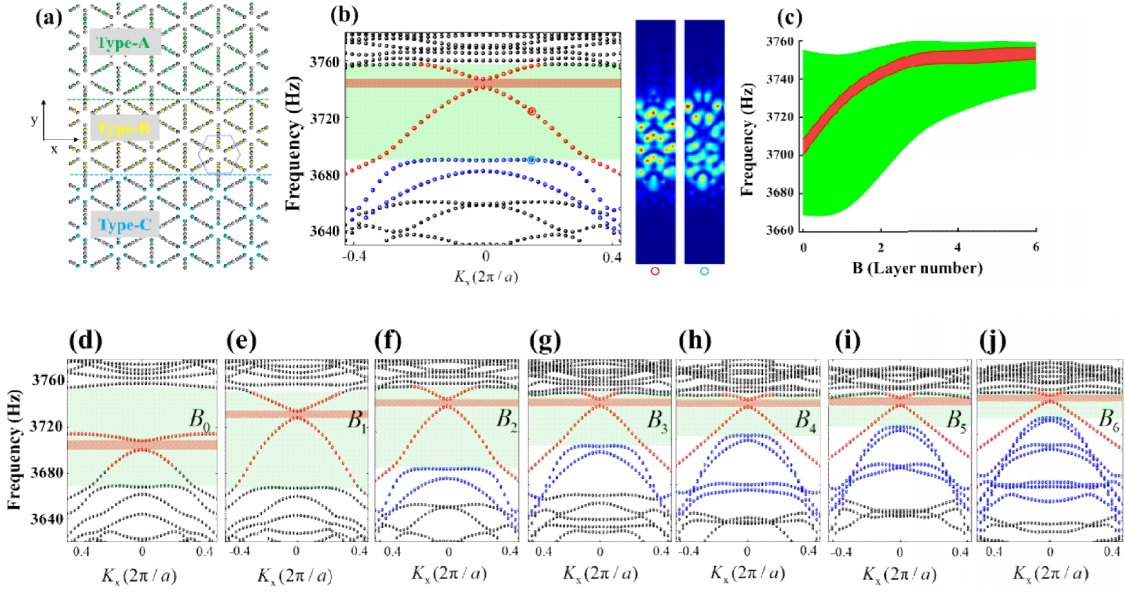

Fig.3 Pseudo-spin locking waveguides. (a) The schematic diagram illustrates a three-layer structure, A|$B_x$|C, where Type-$B_x$ is sandwiched between Type-A and Type-C. Here, $x$ denotes the width of the middle layer B. (b) Dispersion curves of A|$B_2$|C structure, where black, blue and red dots represent bulk, non-topological and topological modes, respectively. The amplitude distribution of topological and non-topological waveguide modes are marked with red and blue circles, respectively. (c) The frequency range of topological states shown by the green shading in (b) is depicted as a function of $x$, where the red region indicates the small gap between the two edge states. (d)-(j) Dispersion curves of A|$B_x$|C for $x = 0$-$6$, where the black dots represent the bulk states, the red dots represent the topological pseudo-spin locking states, and blue dots represent the non-topological modes shifting into the bandgap from the bulk states.

To demonstrate the advantages of pseudo-spin locking waveguide for energy transmission, we established a straight waveguide to observe the process, as shown in Fig.4. The impedance $Z_{air} = \rho_{air} c_{air}$ is set at the boundary around the structure, where $\rho_{air}$ and $c_{air}$ are the air density and the speed of sound in air, respectively. A hard boundary condition is set at the upper end due to the acrylic plates, and the impedance at the intersection of the air column and the liquid surface is set as $Z_{water} = \rho_{water} c_{water}$, where $\rho_{water}$ and $c_{water}$ are the water density and the speed of sound in water, respectively. In order to simulate the impedance tube used in the experimental part, the excitation is chosen to be an acoustic plane wave radiation and placed at the left boundary.

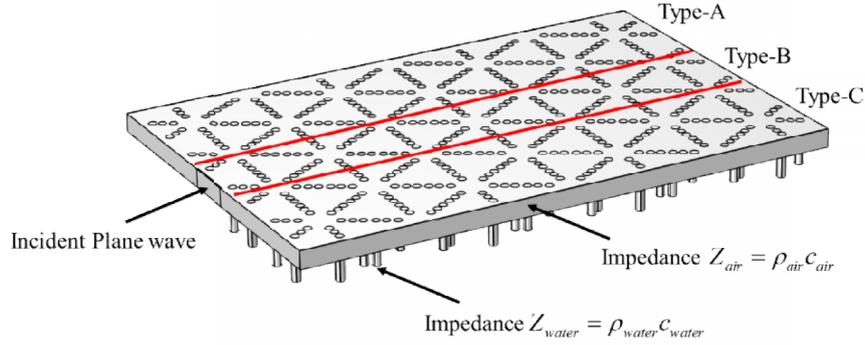

Fig.4 Schematic diagram of the pseudo-spin locking waveguide. The structure consists of Type-A, Type-$B_1$ and Type-C phononic crystals, with the red line indicating the interface between different phononic crystals. Channel height $H_c$ is 15 mm. The height of the air column is $h_d = 20$. A plane wave radiation condition is applied to the left end. A matching impedance $Z_{air} = \rho_{air} c_{air}$ is applied on the bottom and side boundaries of the channel. A matching impedance $Z_{water} = \rho_{water} c_{water}$ is set at the interface between air and water.

A plane wave with pressure of 1 Pa is set at the left end. The pressure amplitude distributions inside the three waveguides are calculated and plotted, the results are shown in Fig.5. Fig.5(a)-(c) show that the acoustic wave decays exponentially in the A and C domains and transmits though the entire B domain. Red stars represent the plane wave radiation source with pressure of 1 Pa. The energy value along the red dashed line of the waveguide is obtained through simulation. As illustrated in Fig.5(d), the total transmitted energy increases with the increase of $x$. The interface states of the heterostructure are superimposed on the bulk state of the intermediate domain $B_x$, which enhances the high energy transport capacity of the wide waveguide.

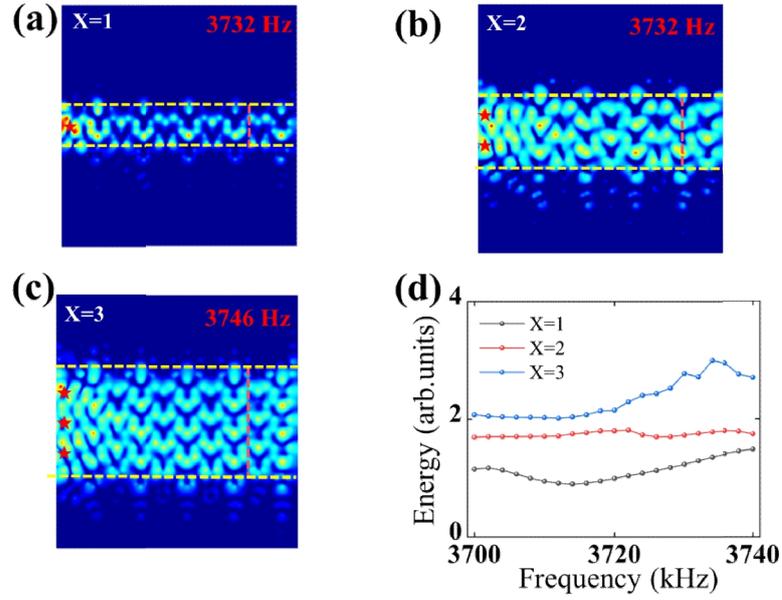

Fig.5. (a) Pressure field of A|B$_1$|C structure at frequency 3732Hz. (b) Pressure field of A|B$_2$|C structure at frequency 3732Hz. (c) Pressure field of A|B$_3$|C structure at frequency 3746Hz. Red stars represent the plane wave excitation sources. (d) The relationship between transmitted energy and the frequency of excitation sources.

A hexagonal array of acoustic sources are set to stimulate counterclockwise (clockwise) energy flow, as shown in Fig.6(a). A broad A|B$_2$|C waveguide is constructed to achieve unidirectional propagation. The spin-locking property of the unidirectional upward (downward) propagation of a topological waveguide, excited by counterclockwise (clockwise) rotational energy, is illustrated in Fig 6(b-c), respectively. Unlike traditional helical edge states, the width of the helical waveguide discussed here can be controlled by change the distribution of the liquid in tubes, allowing large-flux energy transmission.

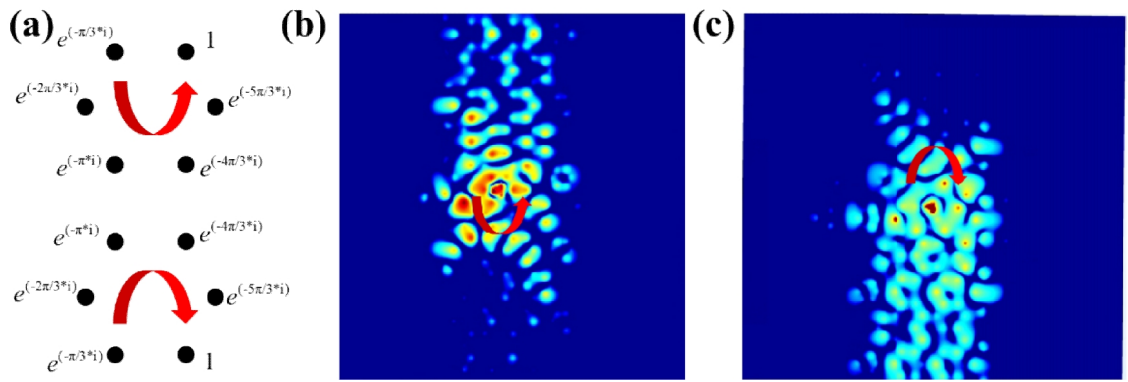

Fig.6 Spin-locked unidirectional transport by using the quantum spin Hall effect. (a) Schematic diagram of acoustic energy excitation. (b,c) Acoustic pressure fields of unidirectional transport excited by counterclockwise and clockwise rotating energy flows, respectively.

Although conventional spin Hall interface states are robust to defects, it is largely dependent on the location and type of defects. Introducing larger defects in conventional spin Hall systems reduces Berry curvature localization and enhances inter-valley mixing, potentially disrupting energy transfer in edge states. For A-C interface states, different defect levels are designed, as shown in Fig 7. In conventional edge state path transmission, a unit-cell lattice defect typically exerts minimal influence, as illustrated in Fig 7(a). However, when defects are designed to be on the scale of two wavelengths, significant backscattering occurs, which impedes energy transmission, as shown in Fig 7(b).

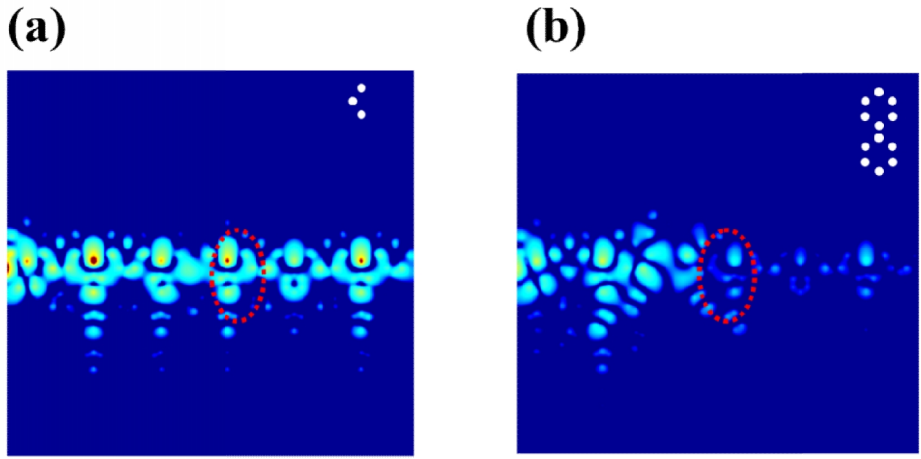

Fig.7 Validation of the robustness of conventional A-C spin Hall model topological waveguides. (a) Simulation of topological waveguide energy propagation for a unit-cell lattice defect. (b) Simulation of topological waveguide energy propagation for two unit-cell lattice defects. The white points at the top-right corner indicates the defect shape and the red circle indicates the location of the defect.

Unlike conventional quantum spin Hall systems, pseudo-spin locking waveguide proposed in this work exhibits exceptional suppression to backscattering, even the large-size defect emerges. Because the energy amplitude is distributed across the entire broad path, the system's transmission robustness is enhanced by its width, which leads that the defects have a significantly reduced effect on the waveguide. To assess how the defect's size affects the system's energy propagation, we set a wavelength-size defect, created by adding the water to remove the air columns, as shown in Fig 8(a). The arrangement of defects with two transverse wavelength dimensions and two longitudinal wavelength dimensions are illustrated in Fig.8(b) and Fig.8(c), respectively. To further demonstrate the strong robustness of the large waveguide system against defects, a larger defect depicted in Fig.8(d) is set. Despite some energy diffusion at the defect location, the system can almost fully transmit the energy. The results indicate that pseudo-spin locking waveguide can withstand larger defects than conventional topological edge states. This research phenomenon will bring more application prospects, such as focusing, beam spreading and wide path sensing.

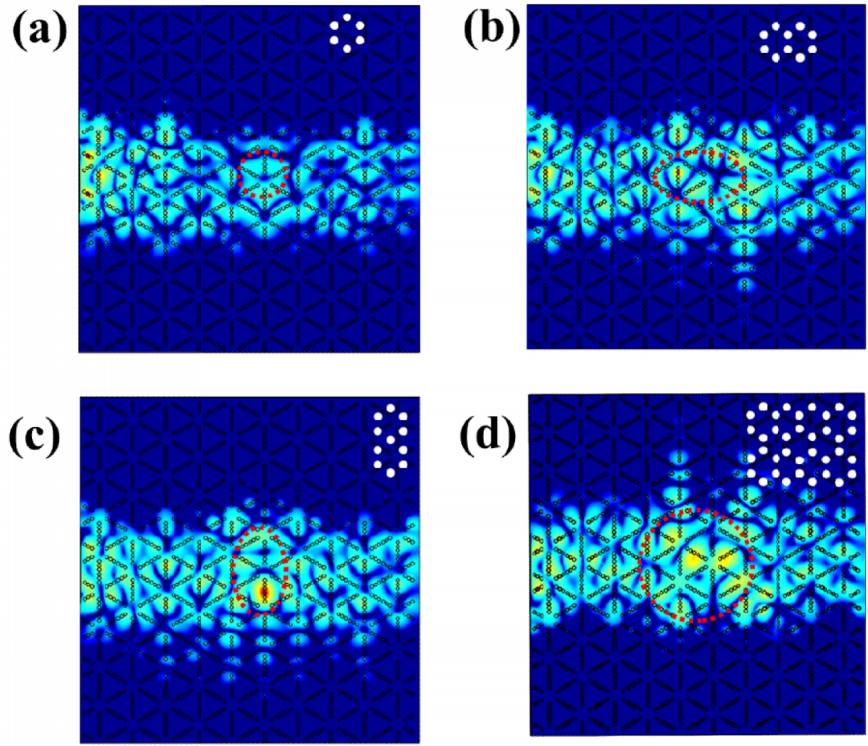

Fig.8 Robustness verification of pseudo-spin locking waveguides. (a) - (d) Simulation results the acoustic wave transmission of pseudo-spin locking A|B$_3$|C waveguide with cavity defects. From (a) to (d), the dimensions of the cavity defects increase along the transverse and the longitudinal direction (see inset schematic). The white points at the top-right corner indicates the cell defect shape and the red circle indicates the location of the defect.

Acoustic energy concentration characteristics of pseudo-spin locking waveguide are explored. Fig.9 illustrates the stepwise acoustic energy transmission of topological energy concentrator. At a third of the path length of the waveguide, three Type-B domains change to a Type-B domain, and then change back to three Type-B domains at a third of the path length. Red stars represent the acoustic excitation source with pressure of 1 Pa. The simulation results indicate that the wave's energy intensity become higher at the narrow part.

To quantitatively assess the enhancement of energy intensity, the energy intensity distributions along the dashed lines are present in Fig. 9(b) for three parts. Along the propagation path, the energy of the $B_1$ channel, depicted by red dashed lines, is significantly greater than that of the two wider $B_3$ channels on the left and right sides, shown in purple and green, respectively. This indicates a strong energy concentration in the middle narrow channel, which can be tuned by adjusting the width of the transmission path.

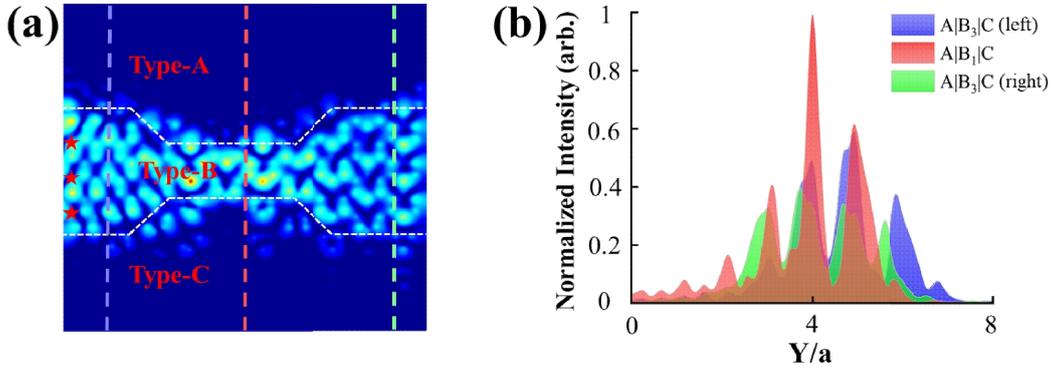

Fig.9 Topological energy concentrator. (a) The sound pressure field simulated at 3740 Hz with the width of the intermediate domain B varying from $x = 3$ to $x = 1$ and back to $x = 3$. Fields A, B, and C are divided by white dashed lines. The red stars represent the acoustic excitation source. (b) The simulated intensity profiles for the points at three dashed lines in (a).

## 4. Acoustic energy intensity amplification based on pseudo-spin locking waveguide and topological rainbow trappings

For topological edge states, by controlling structural parameters, topological rainbow trapping can be realized conveniently. We can simply and easily achieve the gradient distribution by adding water to the hole to change the heights of air columns. Here, the cut-off frequency at which the group velocity approaching zero is directly related to the height of the air column in the unit cell, which is crucial to forming topological rainbow trappings. To illustrate this, we designed the topological rainbow device by varying the heights of air columns in the intermediate domain along the x-direction. Fig. 10(a) shows a schematic of a device consisting of a pseudo-spin locking waveguide A|B$_1$|C, where the yellow line delineating the extent of the B$_1$ domain. The device features a gradient phononic crystal structure composed of 9 unit cells (where Type-A and C have 4 unit cells, respectively) along the y-axis and 10 unit cells along the x-axis. The 10 cells along the x-axis are organized into 5 groups, each of which contains 2 unit cells with identical air column heights. The inset image in Fig. 10(a) presents a cross-section of the B$_1$ domain of the device from the z-plane direction, illustrating that the air column height of the B$_1$ domain gradually increases along the +x direction. The grey circles indicate that none air column exists, while the green, yellow, and blue circles denote various locations of the air column, forming unit cells with three distinct topological phases. The air column height in group $t$ is denoted as $h_d$, which increases linearly with the increase of $t$. Here $h_{d,t} = h_l + (t-1)*h_z$, where $h_l$ is the initial depth and $(t-1)*h_z$ is the change in height. In the example, the step size is $h_z = 0.2\text{mm}$, the initial depth is $h_l = 20\text{mm}$.

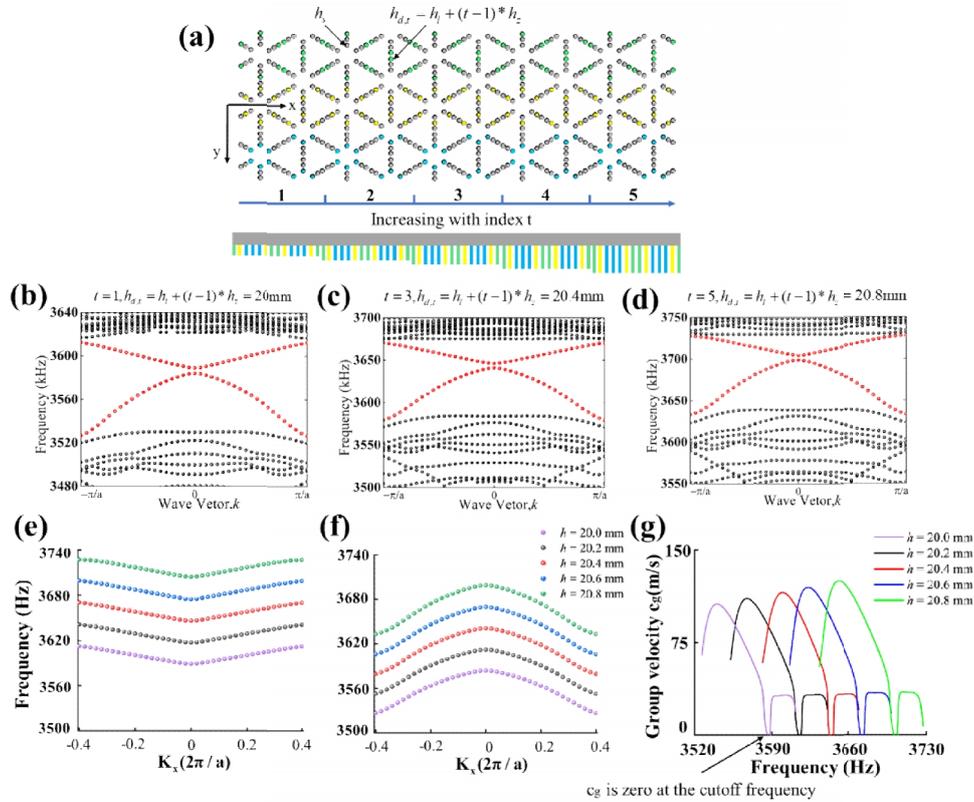

Fig.10 Variation of dispersion curves of gradient phonon crystals based on spin hall effects. (a) Schematic diagram of a gradient phonon crystal with A|B$_1$|C supercell. As the height index '$t$'

increases, the air column height $h_{d,t}$ increases linearly, starting at $h_{d,1}$ = 20.2mm for the first group. The inset model represents a profile in the *z*-direction, illustrating a gradual increase in air column height along the *x*-direction. (b-d）Dispersion relations for group 1, group 3, and group 5 A|B$_1$|C supercells in gradient phononic crystals. (e-f) The up and down edge states for air columns with heights $h_d$ ranging from 20 mm to 20.8 mm. As $h_d$ increases, the interfacial state frequency gradually rises. (g) Group velocity curves for different height $h_d$. The maximum group velocity increases with the increase of $h_d$, and the cutoff frequency $f_{cut}$ increases.

Then, we examine the frequency and group velocity in the supercell dispersion relationship indexed by *t*. Fig.10(b-d) present the dispersion curves for the three supercells at *t* = 1, 3, and 5, while Fig 10(e-f) show the up and down edge states for supercells with *t* = 1-5. The results reveal that the frequency of the edge states rises as *t* increases. Fig.10(g) show the group velocity profiles for changes in air column height ('*t*' varies from 1 to 5). The value of zero of the group velocity curves corresponds the cut-off frequency of the edge state, namely the energy can no longer transmit. As height $h_d$ increases, both the maximum group velocity and the cut-off frequency $f_{cut}$ increase. It is noted that the group velocity of edge states is lower than that of the acoustic waves in air (343 m/s), which can effectively delay acoustic wave transmissions.

Then, the pseudo-spin locking waveguide is designed by using the gradient phononic crystals, allowing for spatial modulation of the edge mode group's velocity. Here, a plane wave of 1 Pa is used to be acoustic energy source, covering the frequency range of 3600 to 3800 Hz. Fig.11 illustrates the acoustic wave responses at four selected frequencies: 3664, 3696, 3725, and 3746 kHz, clearly showing the accumulation of edge mode energy, as shown in Fig.11(a-d). As the acoustic wave of the interface mode propagates, the group velocity gradually decreases to zero, which lead the energy accumulation at distinct locations along the transmission path. It is shown that the accumulation of pseudo-spin locking waveguide energy at different frequencies, forming acoustic rainbow trapping. As the frequency decreases, the point where the group velocity is zero moves forward, which means that the wave can travel further along the *x*-direction at lower frequencies.

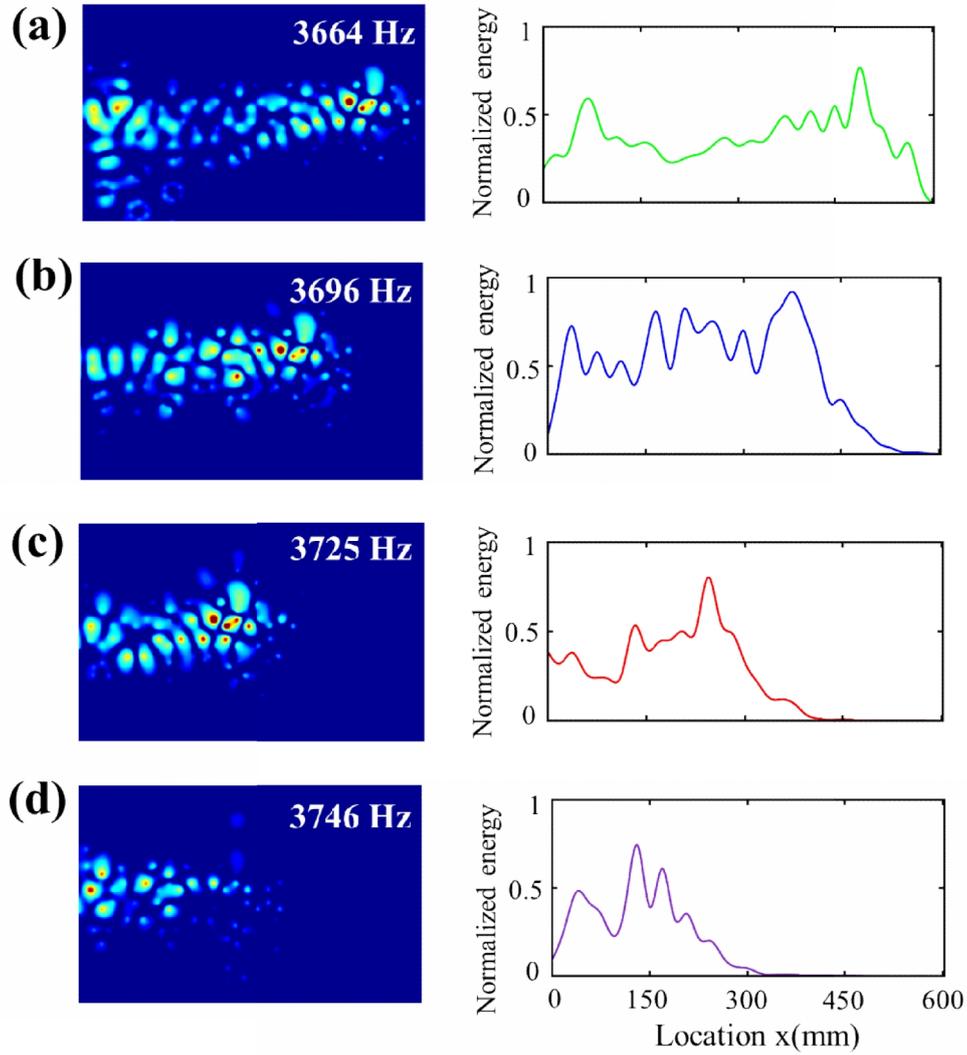

Fig.11 The distribution of acoustic energy fields in pseudo-spin locking waveguide based on gradient phonon crystals. (a-d) The energy distribution in the $z = 35$ mm plane is analyzed at frequencies 3664, 3696, 3725, and 3746 Hz. The normalized wave energy distribution along the acoustic path is extracted from the simulation results, showing that waves of varying frequencies cease propagating at different $x$ positions, resulting in concentrated energy.

For the acoustic energy transmission in pseudo-spin locking waveguides, it is observed that the intensity of the transmitted energy can be modified by altering the width of the transmission path. Fig.12(a) and (b) display the energy diagram of our proposed pseudo-spin locking waveguide systems, featuring intermediate domains $B_1$ and $B_3$, respectively. Fig.12(c) shows the energy distribution of the topological energy concentrator. It is shown that higher energy intensity can be obtained by using topological energy concentrators. In the following, acoustic energy intensities are compared for four kinds of waveguide paths. The energy intensity results are extracted from the red line position in the simulation results, where red star represent the plane wave radiation source of 1Pa, and the results are shown in Fig.12(d). Clearly, the energy obtained by the topology concentrator surpasses that of the other three cases. The large-area topology concentrator amplifies energy by nearly 2.5 times compared to the A|$B_1$|C waveguide transmission. Compared to energy transmission without a path, the large-area topology concentrator amplifies the energy intensity by 167 times.

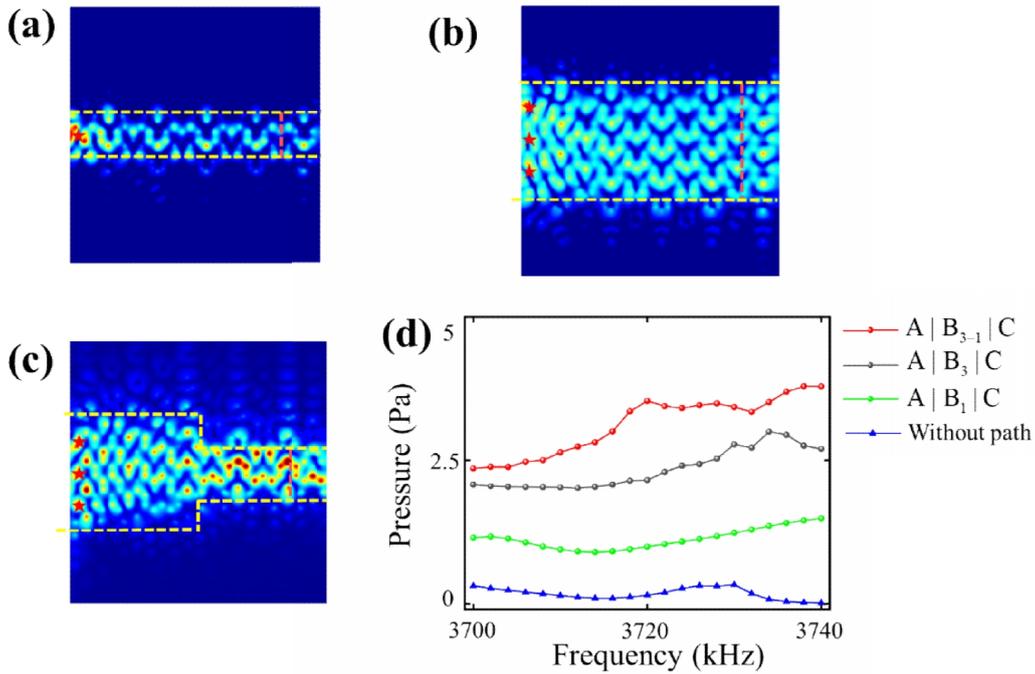

Fig.12 Energy accumulation effect of the pseudo-spin locking waveguide. (a) and (b) show energy distributions of pseudo-spin locking waveguides with types of A|$B_1$|C and A|$B_3$|C, respectively. (c) The energy distributions of topological energy concentrator. (f) The relationship between acoustic pressure and frequency at the red line position for four types of paths.

As we know, when the group velocity approaches zero, the rainbow trapping lead to energy concentration and the acoustic wave pressure is amplified. Furthermore, our tunable system facilitates the coupling of a pseudo-spin locking waveguide topological concentrator and a rainbow trapping. To verify the energy concentration properties, three structures are designed. Fig. 13(a) illustrates the topological rainbow transmission with the A|$B_1$|C, Fig. 13(b) depicts the energy transmission using the

coupling of a pseudo-spin locking waveguide A|B$_{3-1}$|C and a topological rainbow trapping, and the results for coupling of a waveguide A|B$_{3-1}$|C and a topological rainbow trapping are shown in Fig.13(c). It can be seen that effects of the energy concentration of the coupled energy concentrator are obvious. The energy intensity results are extracted from the red line position, where red stars represent excitation sources of 1 Pa, and the energy intensity results are shown in Fig. 13(d).

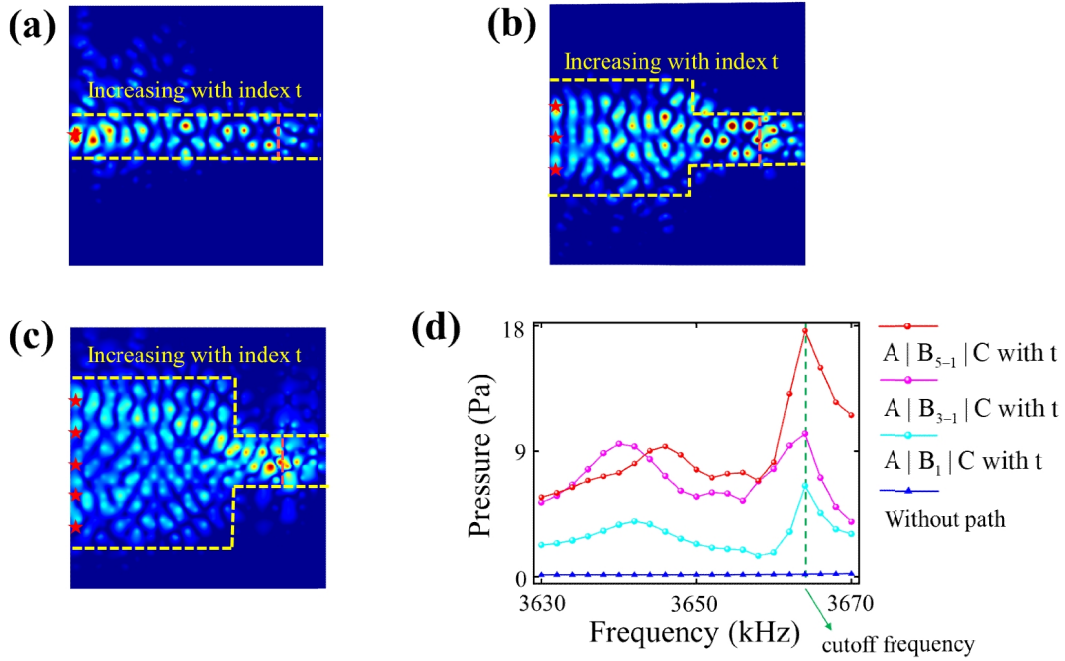

Fig.13. (a) Rainbow energy distributions of the pseudo-spin locking waveguide with type of A|B$_1$|C. (b) The energy distribution of a coupled energy concentrator based on A|B$_3$|C pseudo-spin locking waveguide and a topological rainbow trapping. (c) The energy distribution of a coupled energy concentrator based on A|B$_5$|C pseudo-spin locking waveguide and a topological rainbow trapping. (d) Comparison of the energy intensity for the three devices.

The results show that the incident acoustic pressure can be amplified by 6.5 times through the topological rainbow trapping. Besides, for a coupled energy concentrator based on A|B$_5$|C pseudo-spin locking waveguide and topological rainbow trapping, the energy can be amplified by 17.4 times, which is nearly 770 times higher than that without topological path. Moreover, the results show that altering the width of the topological concentrator can control the level of energy accumulation. In the work, the control of energy intensity amplification and the positioning of energy gathering can by realized conveniently by using the tunable system, which paves the way for advanced energy transmission and control in both acoustics and optics.

## 5. Conclusion

In this work, tunable acoustic energy concentrations is investigated based on pseudo-spin locking waveguides and topological rainbow trappings. A tunable pseudo-spin locking is proposed, the width of the waveguide can be tuned conveniently by adjusting the liquid distribution in tubes, and the broad acoustic energy transport and spin-locked one-way transport effects are verified. The results show that acoustic wave transports based on pseudo-spin locking waveguides are more robust obviously to structure defects than conventional topological edge-state waveguides. Besides, topological rainbow trappings are realized by adjusting the distribution of liquid in tubes. Based on those, the coupling of the pseudo-spin locking waveguides and the topological rainbow trappings are investigated. The results show that high acoustic energy concentrations can be obtained conveniently by using the coupling energy concentrator based on the pseudo-spin locking waveguides and the topological rainbow trappings. The results provide a novel effective method to concentrate acoustic wave energy, which is important in the application field of acoustic sensings and microfludics.

The according experimental verifications will be presented in the near future.

## Appendix A. Adjustment of the ceiling height to change the frequency of the pseudo-spin locking state

The system's tunability allows for the energy transmission at various frequencies, significantly expanding the range of potential applications. As shown in Fig.A1, adjusting the height $H_c$ of the air wall enables easy modification of change the frequency of the pseudo-spin locking state. The capability to alter the frequency of the pseudo-spin locking state enhances the system's versatility and adaptability.

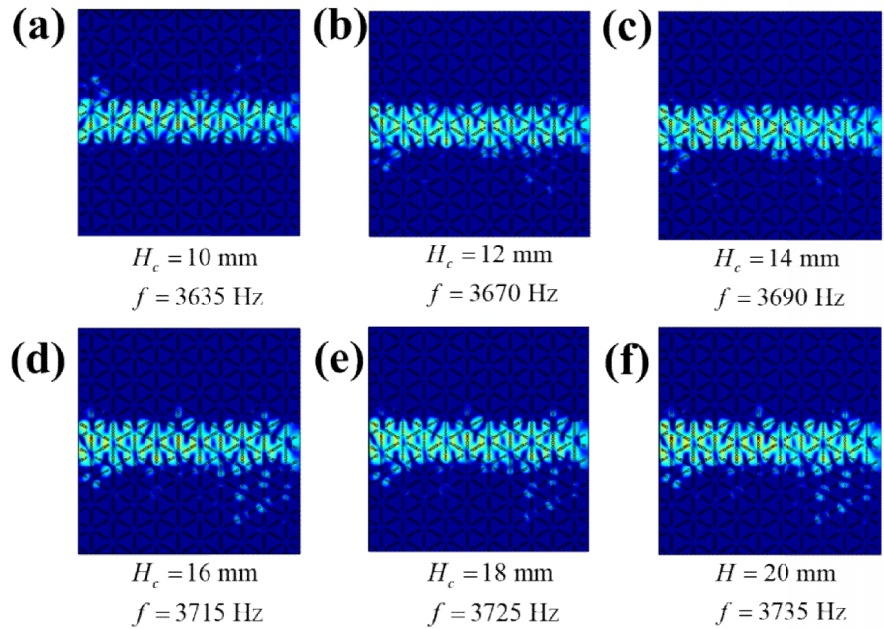

Fig. A1 Variation of the frequency of of the pseudo-spin locking state by changing the height of the ceiling. (a) At $H_C$ = 10 mm, the frequency is 3635 Hz. (b) At $H_C$ = 12 mm, the frequency is 3670 Hz. (c) At $H_C$ = 14 mm, the frequency is 3690 Hz. (d) At $H_C$ = 16 mm, the frequency is 3715 Hz. (e) At $H_C$ = 18 mm, the frequency is 3725 Hz. (f) The simulated pressure field with a frequency of 3735 Hz when $H_C$ =20 mm.